\documentclass[twocolumn,superscriptaddress,preprintnumbers,amsmath,amssymb,prd]{revtex4-2}

\usepackage{graphicx}% Include figure files
\usepackage{dcolumn}% Align table columns on decimal point
\usepackage{bm}% bold math
\usepackage{indentfirst} 
\usepackage{color}
\usepackage{CJK}
\usepackage{booktabs}
\usepackage{float}

\usepackage{ulem}
\usepackage[colorlinks,linkcolor=blue,urlcolor=blue,citecolor=blue]{hyperref}

\usepackage{tikz,xcolor,hyperref}

\definecolor{lime}{HTML}{A6CE39}
\DeclareRobustCommand{\orcidicon}{
	\begin{tikzpicture}
	\draw[lime, fill=lime] (0,0) 
	circle [radius=0.16] 
	node[white] {{\fontfamily{qag}\selectfont \tiny ID}};
	\draw[white, fill=white] (-0.0625,0.095) 
	circle [radius=0.007];
	\end{tikzpicture}
	\hspace{-2mm}
}
\foreach \x in {A, ..., Z}{%
	\expandafter\xdef\csname orcid\x\endcsname{\noexpand\href{https://orcid.org/\csname orcidauthor\x\endcsname}{\noexpand\orcidicon}}
}
\foreach \x in {A, ..., Z}{%
	\expandafter\xdef\csname orcid\x\endcsname{\noexpand\href{https://orcid.org/\csname orcidauthor\x\endcsname}{\noexpand\orcidicon}}
}

\begin{document}
\begin{CJK*}{UTF8}{gbsn}
%\begin{CJK*}{GBK}{song}

%\preprint{Institute of Modern Physics: 201228001707006: first draft}

\title{$t$+$t$ cluster states in $^{6}$He}
%\title{Integrating Experiment and Theory: Insights into $t$+$t$ Cluster States in $^{6}$He}
%\title{Bridging Theory and Experimental Insights of the $t$+$t$ cluster states in $^{6}$He}



%\author{W.H.Ma et al.}
%\email[]{maweihu@fudan.edu.cn}
%\affiliation{Institute of Modern Physics, Fudan University, Shanghai 200433, People's Republic of China}

\author{W.~H.~Ma(马维虎)\orcidA{}}
%\email[]{maweihu@fudan.edu.cn}
\affiliation{Key Laboratory of Nuclear Physics and Ion-beam Application (MOE), Institute of Modern Physics, Fudan University, Shanghai 200433, People's Republic of China}
%\affiliation{Institute of Modern Physics, Fudan University, Shanghai 200433, People's Republic of China}
\author{D.~Y.~Tao(陶德晔)\orcidG{}}
%\affiliation{Institute of Modern Physics, Fudan University, Shanghai 200433, People's Republic of China}
\affiliation{Key Laboratory of Nuclear Physics and Ion-beam Application (MOE), Institute of Modern Physics, Fudan University, Shanghai 200433, People's Republic of China}
\author{B.~Zhou(周波)\orcidF{}}
\email[]{zhou$_$bo@fudan.edu.cn}
\affiliation{Key Laboratory of Nuclear Physics and Ion-beam Application (MOE), Institute of Modern Physics, Fudan University, Shanghai 200433, People's Republic of China}
\affiliation{Shanghai Research Center for Theoretical Nuclear Physics, NSFC and Fudan University, Shanghai 200438, People's Republic of China}
\author{J.~S.~Wang(王建松)\orcidC{}}
\email[]{wjs@zjhu.edu.cn}
\affiliation{School of Science, Huzhou University, Huzhou 313000, People's Republic of China}
\author{Y. G.~Ma(马余刚)\orcidB{}}
\email[]{mayugang@fudan.edu.cn}
\affiliation{Key Laboratory of Nuclear Physics and Ion-beam Application (MOE), Institute of Modern Physics, Fudan University, Shanghai 200433, People's Republic of China}
\affiliation{Shanghai Research Center for Theoretical Nuclear Physics, NSFC and Fudan University, Shanghai 200438, People's Republic of China}
\author{D.~Q.~Fang(方德清)\orcidD{}}
\affiliation{Key Laboratory of Nuclear Physics and Ion-beam Application (MOE), Institute of Modern Physics, Fudan University, Shanghai 200433, People's Republic of China}
\author{W.~B.~He(何万兵)\orcidE{}}
\affiliation{Key Laboratory of Nuclear Physics and Ion-beam Application (MOE), Institute of Modern Physics, Fudan University, Shanghai 200433, People's Republic of China}
%\affiliation{Institute of Modern Physics, Fudan University, Shanghai 200433, People's Republic of China}
\author{Y.~Y.~Yang(杨彦云)}
\affiliation{Key Laboratory of High Precision Nuclear Spectroscopy, Institute of Modern Physics, Chinese Academy of Science, Lanzhou 730000, People's Republic of China}
\author{J.~B.~Ma(马军兵)}
\affiliation{Key Laboratory of High Precision Nuclear Spectroscopy, Institute of Modern Physics, Chinese Academy of Science, Lanzhou 730000, People's Republic of China}
\author{S.~L.~Jin(金仕纶) }
\affiliation{Key Laboratory of High Precision Nuclear Spectroscopy, Institute of Modern Physics, Chinese Academy of Science, Lanzhou 730000, People's Republic of China}
\author{P.~Ma(马朋) }
\affiliation{Key Laboratory of High Precision Nuclear Spectroscopy, Institute of Modern Physics, Chinese Academy of Science, Lanzhou 730000, People's Republic of China}
\author{J.~X.~Li(李加兴) }
\affiliation{School of Physical Science and Technology, Southwest University, Chongqing 400044, People's Republic of China}
\author{Y.~S.~Song（宋玉收） }
\affiliation{Fundamental Science on Nuclear Safety and Simulation Technology Laboratory, Harbin Engineering University, Harbin 150001,  People's Republic of China}
\author{Q.~Hu(胡强)}
\affiliation{Key Laboratory of High Precision Nuclear Spectroscopy, Institute of Modern Physics, Chinese Academy of Science, Lanzhou 730000, People's Republic of China}
\author{Z.~Bai(白真) }
\affiliation{Key Laboratory of High Precision Nuclear Spectroscopy, Institute of Modern Physics, Chinese Academy of Science, Lanzhou 730000, People's Republic of China}
\author{M.~R.~Huang(黄美容) }
\affiliation{College of Physics and Electronics Information, Inner Mongolia University for Nationalities, Tongliao 028000, People's Republic of China}
\author{X.~Q.~Liu(刘星泉)}
\affiliation{Institute of Nuclear Science and Technology, Sichuan University, Chengdu 610064, People's Republic of China}
\author{Z.~H.~Gao(高志浩) }
\affiliation{Key Laboratory of High Precision Nuclear Spectroscopy, Institute of Modern Physics, Chinese Academy of Science, Lanzhou 730000, People's Republic of China}
\author{F.~F.~Duan(段芳芳)}
\affiliation{Key Laboratory of High Precision Nuclear Spectroscopy, Institute of Modern Physics, Chinese Academy of Science, Lanzhou 730000, People's Republic of China}
\author{S.~Y.~Jin(金树亚) }
\affiliation{Key Laboratory of High Precision Nuclear Spectroscopy, Institute of Modern Physics, Chinese Academy of Science, Lanzhou 730000, People's Republic of China}
\author{S.~W.~Xu(许世伟) }
\affiliation{Key Laboratory of High Precision Nuclear Spectroscopy, Institute of Modern Physics, Chinese Academy of Science, Lanzhou 730000, People's Republic of China}
\author{G.~M.~Yu(余功明) }
\affiliation{School of Physical Science and Technology, Kunming University, Kunming 650214, People's Republic of China}
\author{T.~F.~Wang(王涛峰) }
\affiliation{School of Physics, BeiHang University, Beijing 100083, People's Republic of China}
\author{Q.~Wang(王琦) }
\affiliation{Key Laboratory of High Precision Nuclear Spectroscopy, Institute of Modern Physics, Chinese Academy of Science, Lanzhou 730000, People's Republic of China}

\date{\today}

\begin{abstract}
The study of $t$+$t$ cluster states in $^{6}$He provides valuable insights into exotic nuclear structures and the behavior of fermionic cluster systems. This study shows rich cluster resonant state structures above the threshold, identified by experimental reconstruction and theoretical calculations. The excitation energy spectrum above the $t$+$t$ threshold in $^{6}$He is measured via the fragmentation excitation process during the breakup reaction of $^{9}$Li on a $^{208}$Pb target at an incident energy of 32.7 MeV/nucleon. The resonant states are reconstructed from the final state coincident particles $t$+$t$ using the invariant mass method, while the non-resonant background is estimated using the event mixing method. Two states of energy level peaks at $13.9\pm0.3$ and $15.0\pm0.3$ MeV are observed.  Microscopic cluster model calculations exploring the $t+t$ resonant states in $^6\mathrm{He}$ yield theoretical energy spectra which are then compared with the current experimental results. The calculated reduced width amplitudes (RWA) of the $t+t$ channels further confirm the clustering structure of the identified $t+t$ resonant states.
\end{abstract}


\maketitle

\section{Introduction}
The study of nuclear cluster structures expands our understanding of nuclear physics by revealing complex interactions and collective behaviors that go beyond traditional nuclear models \cite{Oertzen,YeYL,YeYL2,MaWH,MaWH1,MaWH2}. Experimental and theoretical advances \cite{WMa,Horiuchi,HeWB,WangSS,CaoYT} explore cluster dynamics, while applications in relativistic collisions \cite{WangYZ,CaoRX,Ma_NT,ZhangYX,ChenJH} highlight their broader relevance. In the NuPECC Long Range Plan 2024 \cite{nupecc}, it states that tritium and $^{3}$He clustering are expected to supersede the $^{4}$He clustering in neutron-rich or neutron-deficient nuclei, respectively. The exploration of these fermionic cluster systems, along with their potential to form quasi-molecular states upon adding further nucleons, is of significant interest and will pose a major challenge to nuclear physics in the coming decade. Triton clustering has been investigated for several decades \cite{Buck,Stratton,Zhongzhou} and has continued to develop up to the present day \cite{Lombardo,Takayuki}.

$^{6}$He is a neutron-rich isotope and has an exotic nuclear structure. It is well known for its ``halo" nature, where two neutrons orbit a compact core of $^{4}$He. Studying the $t$+$t$ cluster states offers insights into how such loosely-bound and neutron-rich systems form and interact. A long-standing question regarding the distinct nature of the $t$-cluster compared to the $\alpha$-cluster and its role in the formation of nuclei containing $t$-clusters \cite{WMa,Furumoto,Yamada,Satsuka,Kanada}. While $\alpha$-clusters exhibit bosonic behavior, tritons are fermionic, making the contribution of $t$-clusters to nuclear formation a fascinating problem. Elucidating the differences between $t$-clusters and $\alpha$-clusters is expected to offer new insights into nuclear structure. Investigating the structures of $t+\alpha$ \cite{tao2025prc,Tombrello}, $t+2\alpha$ \cite{Kanada,Kanada1,Suhara,Zhou}, $t+\alpha+n+n$ \cite{Furumoto1}, $t+^{6}$He \cite{WMa,Kanada2,Kanada3}, $t+t$ \cite{Povoroznyk}, and $t+t+t$ \cite{Muta}, thereby advancing the cluster formation knowledge, which has garnered significant attention in both experimental and theoretical research over the past decade. Understanding the formation mechanisms of neutron-rich clusters in $^{6}$He, particularly the role of the $t$ cluster, remains an important open problem.
Exploring the fermionic nature of tritons and their impact on nuclear formation, addressing gaps in the study of non-$\alpha$ cluster states. 

Discrepancies are evident in the reported energy level distributions of $^{6}$He across various sources; while the compilation by Ajzemberg-Selove \cite{Ajzenberg} identifies only one low-lying state, Tilley et al. \cite{Tilley} compilation reveals two states capable of decaying into an $\alpha$ particle and two neutrons. Above the decay threshold of 12.31 MeV for $^{6}$He* states into $t+t$ clusters, the number of levels and their energies differ between the two compilations. Additional experiments, such as the observation of a $^{6}$He* state at 18.0 MeV in the $^{6}$Li($^{7}$Li, $^{7}$Be)$^{6}$He  \cite{Yamagata} reaction has yielded findings not included in the aforementioned compilations. The states in the $^{7}$Li(n, d)$^{6}$He \cite{Brady} reaction have found the evidence of excited states at 13.6, 15.4, and 17.7 MeV above the threshold. Resonances at 15.0, 18.0, and 25.5 MeV above the $t+t$ threshold have been studied via the $^{6}$Li($^{7}$Li, $^{7}$Be t)$^{3}$H reaction \cite{Akimune}. The study of $^{6}$He is particularly intriguing due to its behavior as an $\alpha$-particle core with a halo of two neutrons at low excitation energies, contrasting with its structure as two $t+t$ clusters at higher energies. Motivated by these discrepancies,  Povoroznyk et al. \cite{Povoroznyk}, resolved to investigate further three-body reactions, specifically $^{3}$H($\alpha$, $tt$)$^{1}$H. Their high precision observed energy levels of $^{6}$He* are 14.0$\pm$0.4, 16.1$\pm$0.4, and 18.3$\pm$0.2 MeV. with $\Gamma$ widths of 0.7$\pm$0.3, 0.8$\pm$0.4, and 1.1$\pm$0.3 MeV respectively. Nevertheless, discrepancies still exist and have not been fully clarified. The current experimental and theoretical research has made efforts to study the resonant states of $t+t$ cluster structures in $^{6}$He, and the first observation of 17.016 MeV and 19.4 MeV states has been achieved, illustrating the interesting information of energy level structures above the threshold. 

In the realm of theoretical research, the study of the cluster structure of the neutron-rich and unstable $^{6}$He nucleus has long captivated nuclear physicists. Initially, the intriguing "neutron halo" in light nuclei near the drip line, particularly $^{6}$He, motivated the development of sophisticated models \cite{Romero_PRL,Vary_NST}. The extended three-cluster model, which better represents the alpha particle's internal dynamics, emphasizes core breakup effects and the role of $t+t$ clustering \cite{Tanihata, Arai, Csoto}. Advanced computational approaches, such as the microscopic multicluster model and resonating group method, have enhanced our understanding of the structure and dynamics of $^{6}$He, providing deeper insights into neutron-rich nuclei near the drip line \cite{Arai1}. While prior research on the $^{6}$He nucleus has predominantly focused on its ground and low-lying states, particularly its neutron halo structure, the theoretical investigation of resonant states above the $t+t$ threshold has been relatively underexplored. In this study, we extend the application of the Generator Coordinate Method (GCM)  to investigate the $t+t$ cluster states of $^{6}$He in excitation regimes above the threshold, providing new insights into the behavior of the nucleus in these higher-energy states. GCM is a powerful tool in nuclear physics, widely used to describe complex nuclear systems by approximating collective motion through a set of quantum states \cite{Zhou1,Zhou1_view,Zhao}. This method constructs a wave function by integrating over a continuous set of basis states, known as generator coordinates, which correspond to different nuclear configurations. By incorporating collective degrees of freedom such as vibrations and rotations, the GCM is particularly well-suited for studying the structure and reactions of nuclei. 
%This work represents the latest theoretical development in understanding the $t+t$ cluster structure in $^{6}$He.
About forty years ago, various ${^3\mathrm{He}}+{^3\mathrm{He}}$ scattering properties, especially the spin-orbit coupling effects, were analyzed with the system wave functions obtained by GCM~\cite{Baye1981}. 
As a mirror system, exploring $t+t$ clustering states within the GCM framework is indispensable.

This paper is organized as follows. Section II presents the experimental results. In Section III, we describe the framework of the GCM and the reduced width amplitudes (RWA) calculations for $^{6}$He. Section IV provides a detailed discussion, and finally, the conclusions are summarized in Section V.

\section{Experiment}
The experiment was conducted at the Heavy Ion Research Facility in Lanzhou (HIRFL)\cite{WMa}. The experimental setup is depicted in FIG.~\ref{fig1} \cite{WMa1}, which clearly shows an overall configuration of the detector array. A primary $^{12}$C beam, accelerated to an energy of 53.7 MeV per nucleon by the HIRFL, was directed onto a $^{9}$Be target with a thickness of 3038 $\mu$m to generate the secondary beam. This secondary $^{9}$Li beam, with an energy of 32.7 MeV per nucleon, was then separated and purified using a 2153 Î¼m thick aluminum degrader at the Radioactive Ion Beam Line in Lanzhou (RIBLL) \cite{ZSun}. The intensity of the secondary $^{9}$Li beam was approximately $1.1 \times 10^{3}$ particles per second, with a purity exceeding 90$\%$. A self-supported natural lead (Pb) target with a thickness of 526.9 mg/$cm^{2}$ was utilized. Three parallel-plate avalanche counters (PPACs)\cite{PMa}, each with a position resolution better than 1 mm, were placed before the reaction target to determine the position and incident angle of the secondary beam on an event-by-event basis. A zero-degree telescope array was employed to measure the charged fragments. This array consisted of two $\Delta$E detectors (double-sided silicon strip detectors, DSSDs) and an 8$\times$8 CsI(Tl) scintillator array as an E detector, covering an angular range from 0$^{\circ}$ to 10$^{\circ}$. The two DSSDs, with thicknesses of 523 $\mu$m and 527 $\mu$m respectively, had the same sensitive area of 49$\times$49 $mm^{2}$. Each DSSD was divided into 16 strips on both the front and rear sides, with each strip being 3 mm wide and spaced 0.1 mm apart. Each CsI(Tl) scintillator was a frustum of a pyramid, with dimensions of 21$\times$21 $mm^{2}$ at the front end, 23$\times$23 $mm^{2}$ at the back, and a length of 50 mm. These scintillators were coupled with photomultiplier tubes (PMTs). A detailed description of this type of telescope array can also be found in Refs.~\cite{SJin,WMa}. The secondary beams of $^{9}$Li, $^{6}$He, $^{4}$He, and $^{3}$H, produced from the $^{12}$C primary beam, were used to calibrate the telescope. Further details of the experimental setup and measurements are available in our previous publication \cite{WMa,WMa1}.

\begin{figure}[!htb]
\centering
\label{fig1}
\includegraphics[width=0.8\hsize]{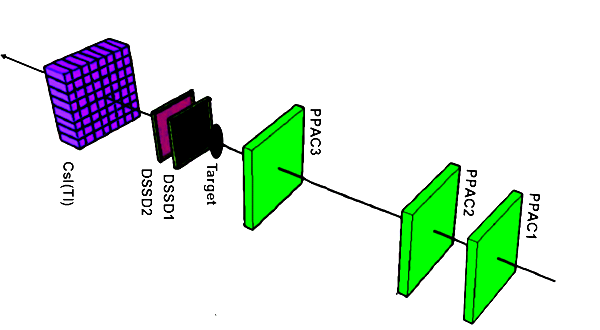} 
\caption{ (Color online) Schematic view of the detector setup \cite{WMa1}.}
\label{fig1}
\end{figure}

\begin{figure}[!htb]
\centering
\label{fig2}
\includegraphics[width=0.9\hsize]{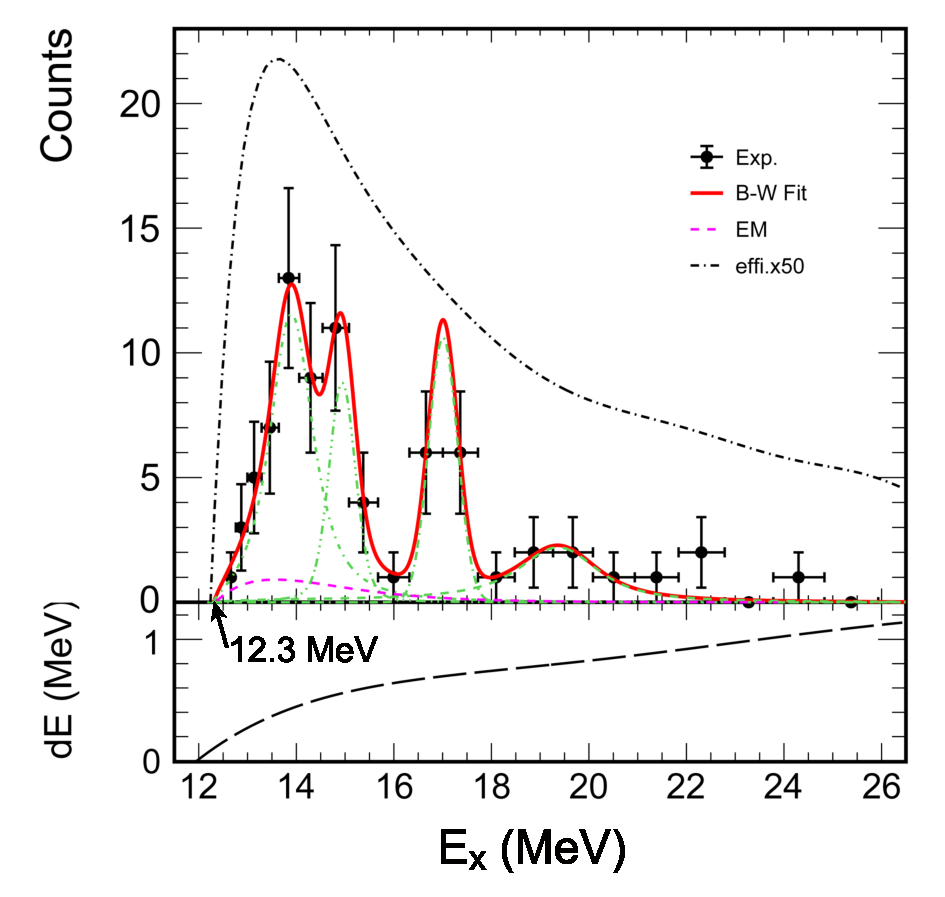} %BWfits_Resolution_final or BWfits_final
\caption{(color online) The $E_{x}$ spectrum, reconstructed from the $t$+$t$ coincident fragments, is fitted with Breit-Wigner-shaped resonance functions (red line) together with the term of nonresonant contribution (pink dashed line marked as EM), taking the resolution (dE) and acceptance efficiency into account. The curve of the acceptance efficiency is scaled by a factor of 50 for better visibility alongside the event counts. }
\label{fig2}
\end{figure}

The experimental configuration employed a high-resolution telescope array for charged particle identification, with the particle identification (PID) plot referenced in the related publication \cite{WMa,WMa1}. Triton particle events, produced via the decay of excited $^{6}$He, were carefully selected based on the telescope array's observations. The total kinetic energy of these triton particles was determined by summing the energy deposited in the double-sided silicon detectors (DSSDs) and the residual energy in the CsI(Tl) scintillator. Additionally, the DSSDs recorded particle trajectories, which were crucial for determining the excited energy ($E_{x}$) of the fragment pairs using the invariant mass method. To further ensure precision, a Monte Carlo simulation was conducted to evaluate the resolution of the excited energy and the geometric detection acceptance, taking into account both the detectors' spatial and energy resolution, the simulation method is the same as Ref. \cite{WMa1}. The same Monte Carlo simulation methodology has been validated and successfully applied in \cite{Freer,ZHYang1,ZHYang2}.  

As shown in FIG. ~\ref{fig2} (top) the detectors' spatial acceptance is $\sim43\%$ at 13.5 MeV of excitation energy. In FIG. ~\ref{fig2} (bottom), the simulation results indicated a mass resolution of approximately 0.4 MeV at an excited energy of 14.0 MeV, increasing to about 0.8 MeV at an excited energy of 20.0 MeV. This trend indicates that higher precision can be achieved at excitation energies near the $t+t$ threshold. The energy resolution curve is used to define bins with unequal spacing (Non-Uniform Binning), allowing for a statistical analysis of the excitation energy spectrum. Using an energy resolution curve to define bins with unequal spacing for the analysis of excitation energy spectra offers key advantages \cite{QYang,Urban}. This approach enhances accuracy by matching bin widths to the detector's energy resolution, improves the signal-to-noise ratio in low-resolution regions by reducing statistical fluctuations to enhance data clarity, and optimizes data representation, ensuring efficient use of statistics without losing key spectral details.

The fragmentation excitation process refers to a nuclear reaction mechanism in which a high-energy projectile, such as a heavy ion, collides with a target nucleus, causing the projectile to break up into smaller fragments. These fragments often populate excited states, with their energy levels determined by the interaction dynamics and detected by decayed final state particles \cite{GGAdamian}. The excitation energy spectrum of $^{6}$He above the $t+t$ threshold is measured through the fragmentation and subsequent decay of a $^{9}$Li projectile on a $^{208}$Pb target. Using the invariant mass method, this experiment explores the excitation energy distribution of $^{6}$He in states above the triton binding threshold. In this reaction, the projectile nucleus undergoes fragmentation, producing lighter fragments such as $^{6}$He in excited states, which subsequently decay into $t+t$.

Several reaction channels are possible in the breakup reaction of \( ^9\text{Li} \) on a \( ^{208}\text{Pb} \) target, which leads to the production of an excited state of \( ^6\text{He} \). One potential reaction channel is the direct breakup, where the \( ^9\text{Li} \) nucleus collides with \( ^{208}\text{Pb} \) and breaks up into \( ^6\text{He} \) and other light fragments. Typical reaction paths include \( ^{208}\text{Pb}(^9\text{Li}, ^6\text{He} + t ) \), \( ^{208}\text{Pb}(^9\text{Li}, ^6\text{He} + n + d ) \), or \( ^{208}\text{Pb}(^9\text{Li}, ^6\text{He} + 2n + p ) \). Another possible channel is the coupled channel reaction, where interactions between the heavy \( ^{208}\text{Pb} \) nucleus and \( ^9\text{Li} \) affect the breakup process, leading to the production of an excited \( ^6\text{He} \). Additionally, transfer reactions may occur, where nucleons from \( ^9\text{Li} \) are transferred to the \( ^{208}\text{Pb} \) nucleus, leaving \( ^6\text{He} \) as an emission product. The reaction channel \( ^{208}\text{Pb}(^9\text{Li}, ^6\text{He} + t ) \) suggests the decay of \( ^9\text{Li} \) into three tritons, with \( ^6\text{He} \rightarrow t + t \). However, due to the limited detection efficiency and low reaction cross-section for \( ^9\text{Li} \) decay into three tritons in the current experimental setup, no coincident events involving three tritons have been observed. It should be noted that if $^9\text{Li}$ decays via a cascade process ($^9\text{Li} \rightarrow {^6\text{He}} + t$ followed by ${^6\text{He}} \rightarrow t + t$), only the two tritons originating from the ${^6\text{He}}$ decay are considered as signal. In contrast, any pair consisting of one triton from the ${^6\text{He}}$ decay and one from the initial step, as well as any pair from a direct $^9\text{Li} \rightarrow t + t + t$ decay, is treated as part of the non-resonant background and is estimated using the event-mixing method \cite{ZHYang}.

Understanding the excitation energy levels and the widths of excited states in light nuclei is essential for validating nuclear models and advancing research in astrophysics \cite{Povoroznyk}. To analyze this, we employ a composition function of double Breit-Wigner (BW) resonance function incorporating a background term, which is convoluted with the detection resolution, while also accounting for the acceptance efficiency. This method is applied to fit the first two peaks in the $E_{x}$ spectrum, as shown in FIG. ~\ref{fig2}. We use double BW functions overlapped on the fitted line shapes from the first two peaks to profile the third and fourth peaks. However, due to an insufficient number of data points, no reliable fitting parameters could be obtained for these higher-order peaks. The background contribution from non-resonant events is estimated using the "event-mixing"  (EM) technique \cite{ZHYang,WMa}. This approach, combining resonance modeling with an event-mixing technique for background estimation, ensures a more accurate description of the excitation spectrum.

\begin{table}[ht]
\centering
\caption{Experimental excitation energy $E_{x}$ and width $\Gamma$, fitted by Least $\chi^{2}$ method with non-uniform binning.}
%\begin{tabular}{|c|c|c|c|c|}
\begin{tabular}{ccccc}
\hline
\hline
\textbf{Peak No.} & \textbf{$E_{x}$ (MeV)} & \textbf{$\Gamma$ (MeV)} & \textbf{$\chi^{2}$/NDF} \\ \hline
1      & 13.93$\pm$0.26      & 0.89$\pm$0.94      & 0.61/2      \\ %\hline
2      & 14.95$\pm$0.30      & 0.15$\pm$1.21      & 0.61/2    \\ %\hline
\hline
\end{tabular}
\label{tab:example}
\end{table}

The results of the Least $\chi^{2}$ fit with non-uniform binning are presented in TABLE I. The uncertainties for the peak parameters are the errors obtained from the least-squares fit applied to the reconstructed invariant mass spectrum. These errors, derived from the covariance matrix during fitting, represent statistical uncertainties under the assumption that the model is correct and that data errors are known. In our fitting process, the measured peak shape is deconvolved with the estimated energy resolution to extract the intrinsic peak parameters. Moreover, fitting four peaks requires 12 parameters, plus one background parameter, totaling 13 parameters. This is indeed challenging with limited data, and appropriate binning is essential to ensure a reliable fit. We adopted non-uniform binning based on the energy resolution curve, which optimizes the signal-to-noise ratio by matching bin widths to the resolution and minimizing statistical fluctuations. This approach enhances data clarity and determines a stable profile. However, in high-excitation energy regions with wide bins, the effective number of bins is reduced, thus lowering the fit's degrees of freedom and making fitting parameters less reliable.

The energy levels fit demonstrates a comparable precision to the study conducted by O. M. Povoroznyk et al., which provided precise measurements of the energy levels and widths of the \(^{6}\text{He}\) state \cite{Povoroznyk}, surpassing earlier research \cite{Ajzenberg,Tilley,Yamagata,Brady,Akimune} that often reported broader and less-defined energy levels. The first energy level at $13.9\pm0.3$ MeV shows strong agreement with the findings of M. Povoroznyk et al., while the second level at $15.0\pm0.3$ MeV aligns with earlier studies. Due to its limited statistics, two candidate peaks around 16.9 MeV and 19.5 MeV need to be further determined. In contrast, the levels at 16.1 and 18.3 MeV identified by Povoroznyk et al. are not observed in this work. This study achieves the same precision for the lower two energy levels, while Povoroznyk et al. demonstrate superior accuracy at higher energies. The width of the first energy level is consistent with Povoroznyk et al., though with a larger error margin. Overall, the fitting errors for width are large, and future precise measurements are essential.

\section{Framework of GCM and RWA calculation of $^6$He}

To microscopically describe the $^{6}$He nucleus we adopt the GCM including the $t+t$ and $\alpha+n+n$ configurations.
The wave function can be written as
\begin{equation}
    \Psi^{J\pi}_M = \sum_{i,K,h}c_{i,K,h} P_{MK}^{J\pi} \Phi_{h}^\mathrm{B}(\{\boldsymbol R\}_i),
\label{wf_tot}
\end{equation}
in which $P_{MK}^{J\pi}$ is the angular-momentum and parity projector and the Brink wave functions with different channels are defined by
\begin{align}
    &\Phi^\mathrm{B}_{tt}(\boldsymbol R_1,\boldsymbol R_2)=\mathcal{A}\left\{\Phi^\mathrm{B}_{t}(\bm{R}_1)\Phi^\mathrm{B}_{t}(\bm{R}_2)\right\},   \\
    &\Phi^\mathrm{B}_{\alpha nn}(\boldsymbol R_1,\boldsymbol R_2,\boldsymbol R_3)=\mathcal{A}\left\{\Phi^\mathrm{B}_{\alpha}(\bm{R}_1)\phi_{n}(\bm{R}_2)\phi_{n}(\bm{R}_3)\right\},  \\
    &\Phi^\mathrm{B}_{\alpha}(\boldsymbol R_j)=\mathcal{A}\left\{\phi_n(\bm{R}_j)\phi_n(\bm{R}_j)\phi_p(\bm{R}_j)\phi_p(\bm{R}_j)\right\},   \\
    &\Phi^\mathrm{B}_{t}(\boldsymbol R_j)=\mathcal{A}\left\{\phi_n(\bm{R}_j)\phi_n(\bm{R}_j)\phi_p(\bm{R}_j)\right\},
\end{align}
serve as the basis wave functions.
$\Phi^\mathrm{B}_{tt}$ and $\Phi^\mathrm{B}_{\alpha nn}$ are the Brink wave functions for configurations $t+t$ and $\alpha+n+n$, respectively.
The wave function of the $k$-th nucleon is defined as the Gaussian wave packet
\begin{equation}
    \phi_k(\bm R_j)=\frac{1}{(\pi b^2)^{3/4}}\exp\left[-\frac{1}{2b^2}(\bm{r}_k-\bm{R}_j)^2\right] \chi_k \tau_k ,
\end{equation}
where the oscillator parameter is set as $b=1.46~\mathrm{fm}$.
The coefficients $\{c_{i,K,h}\}$ are determined by solving the Hill-Wheeler equation.  
% parameters
In the present calculation, the Hamiltonian includes kinetic, central N-N, spin-orbit, and Coulomb terms
\begin{equation}
\begin{aligned}
    H&=-\frac{\hbar^2}{2m}\sum_i \nabla_i^2 - T_\mathrm{c.m.}\\
      &+ \sum_{i<j}V_{ij}^\mathrm{NN} + \sum_{i<j}V_{ij}^\mathrm{LS} + \sum_{i<j}V_{ij}^\mathrm{C}.
\label{ham}
\end{aligned}
\end{equation}
The central N-N potential is taken as the Volkov No.2 potential \cite{Volkov1965}
\begin{equation}
    V_{ij}^\mathrm{NN}=\sum_{n=1}^2v_ne^{-\frac{r_{ij}^2}{a_n^2}}(W+BP_\sigma-HP_\tau-MP_{\sigma\tau})_{ij}
\end{equation}
with $a_1={1.01}~\mathrm{fm}$, $a_2={1.8}~\mathrm{fm}$, $v_1={61.14}~\mathrm{MeV}$, $v_2={-60.65}~\mathrm{MeV}$, $W=1-M$, $M=0.6$ and $B=H=0.125$.
The spin-orbit term is taken as the G3RS potential \cite{Tamagaki1968,Yamaguchi1979}
\begin{equation}
    V_{ij}^\mathrm{LS}=v_0(e^{-d_1r_{ij}^2}-e^{-d_2r_{ij}^2})P(^3O)\bm L\cdot\bm S,
\end{equation}
where $P(^3O)$ is the projection operator onto a triplet odd state, the strength $v_0=2000~\mathrm{MeV}$, and the parameters $d_1$ and $d_2$ are taken as $5.0~\mathrm{fm}^{-2}$ and $2.778~\mathrm{fm}^{-2}$, respectively. 
The oscillator parameter and interaction parameters are set to standard values, which are widely used in various cluster methods.

% RWA
The obtained GCM wave functions of ground and excited states can be used to analyze the cluster structures in the nuclei.
By the use of the Laplace expansion method \cite{Chiba2017}, we calculate the reduced-width amplitudes (RWA) defined as \cite{Descouvemont2010,tao2024,tao2025}

\begin{equation}
\begin{aligned}
&y_{j_1\pi_1j_2\pi_2j_{12}}^{J\pi}(a)=\sqrt{\frac{A!}{(1+\delta_{C_1C_2})C_1!C_2!}}\times\\
    &\left<\frac{\delta(r-a)}{r^2}\left[Y_l(\hat{\bm r})\otimes
        \left[\Phi_{C_1}^{j_1\pi_1}\otimes\Phi_{C_2}^{j_2\pi_2}\right]_{j_{12}}\right]_{JM}\middle|\Psi^{J\pi}_M\right>.
\label{eq:rwa}
\end{aligned}
\end{equation}

\begin{figure*}[!htb]
\centering
\label{fig3}
\includegraphics[width=0.90\hsize]{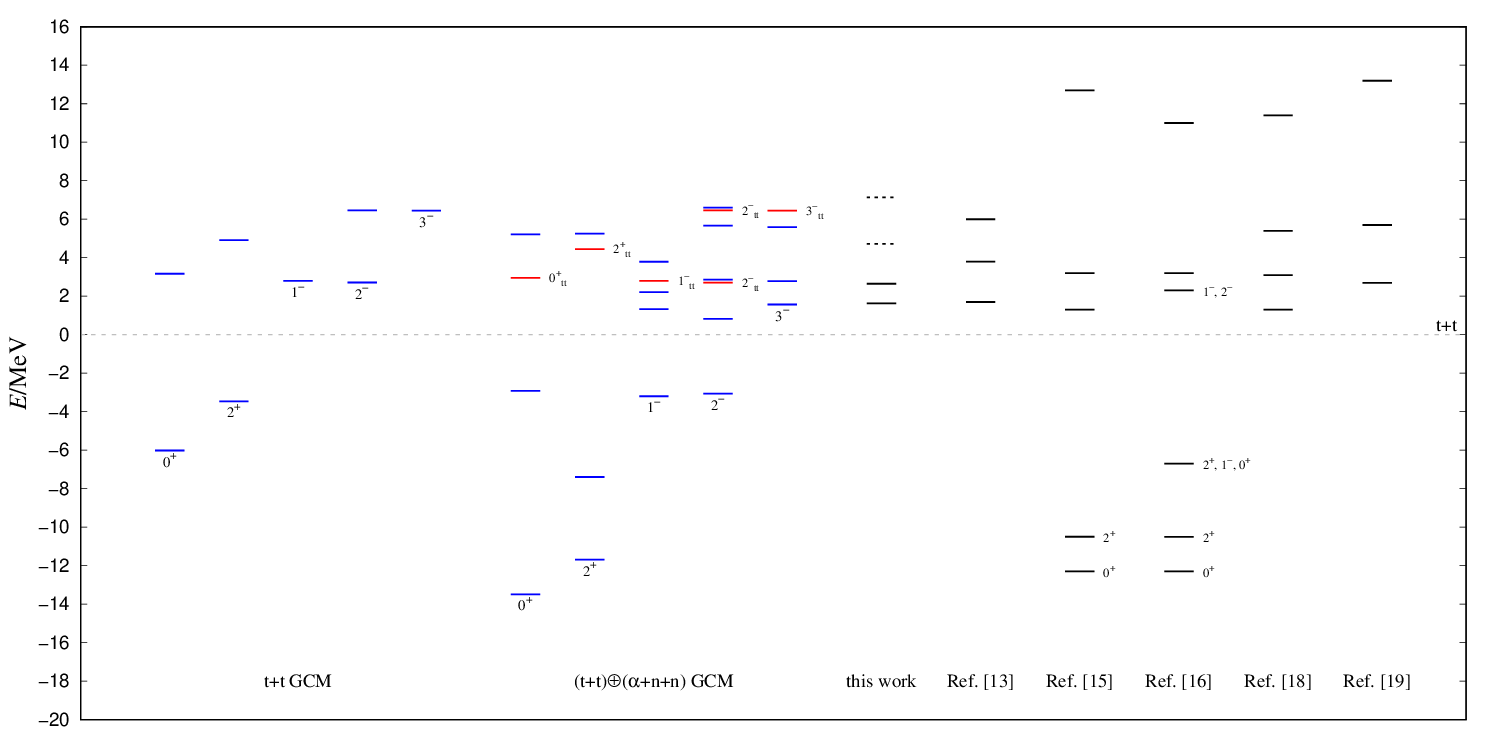}
\caption{(color online) The energy levels of $^{6}$He, calculated using the GCM, are provided for the 0$^{+}$, 2$^{+}$, 1$^{-}$, 2$^{-}$, and 3$^{-}$ states. The energy spectrum obtained from the $t+t$ cluster model is displayed in the left columns. In the middle columns, we present the results from coupling the $t+t$ configuration with the shell-model-like $\alpha+n+n$ configurations and states exhibiting significant $t+t$ structure are highlighted in red. In the right columns, the levels depicted in black represent data from experiments, including this work and previous studies. The dashed lines are the candidates for further confirmation.}
\label{fig3}
\end{figure*}

\section{Discussions}
The energy levels above the $t+t$ threshold exhibit a rich resonance structure, including two states $13.9$ and $15.0$ MeV reported in this study. However, the levels of 16.1 and 18.3 MeV, identified by Povoroznyk et al., are not observed in this work. These discrepancies highlight that, despite the rich resonance structure above the $t+t$ threshold, there is no consistent agreement across all experimental results regarding the excitation energy levels of these states. Furthermore, current measurements, including previous studies, have been unable to accurately determine the spin and parity of the excited states due to experimental accuracy and statistical data limitations. Povoroznyk et al. employed the three-body reaction pathway $^3H(\alpha, tt)^1H$, whereas the current experiment is based on the fragmentation ($^{6}$He) excitation process during the breakup reaction of $^{9}$Li on a $^{208}$Pb target. Due to the effects of incident particle energy, angular momentum matching, nuclear barrier penetration, and selection rules, different reaction mechanisms may lead to variations in the excitation efficiency of specific energy levels. The GCM calculations (FIG. ~\ref{fig3}) suggest that the t+t resonant states of $^{6}$He may exhibit complex quantum structures, yet the absence of spin-parity measurements in the experiment limits a complete characterization of these states. The 16.1 MeV and 18.3 MeV states observed by Povoroznyk et al. may correspond to specific quantum states that were not effectively excited or identified in the current experiment's reaction mechanism or detection setup. Future investigations should focus on spin-parity measurements, higher-statistics experiments, and multi-reaction channel comparisons to further clarify the resonance structure of $^6$He.

To address this, we perform GCM calculations to investigate the quantum state characteristics of the excitation energy spectrum of the $t+t$ cluster structure in $^{6}$He.
By employing the microscopic cluster model and effective nucleon-nucleon interactions, the theoretical results should be independent of any specific reaction channel or detection method.
On the other hand, the GCM calculations suggest the spin and parity of various $t+t$ resonant states, which are crucial for unambiguously identifying these states but have not yet been provided by previous experiments.
These calculations aim to provide guidance and recommendations for more precise measurements in future experiments. 

\begin{figure}[!htb]
\centering
\includegraphics[width=\hsize]{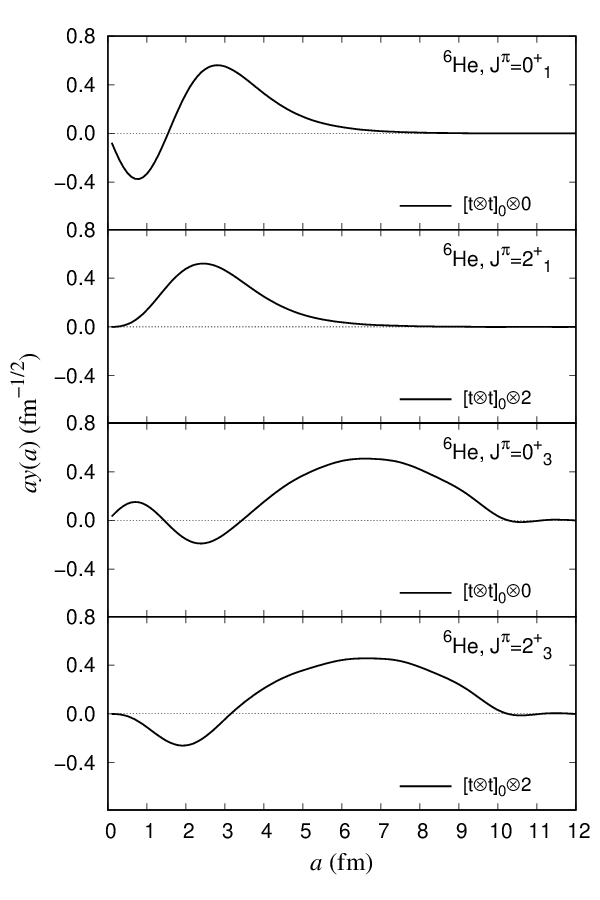}
\caption{The calculated RWA of $t+t$ configuration in the ground state, the first excited state, and positive-parity resonant states with prominent $t+t$ structure.}
\label{fig4}
\end{figure}

\begin{figure}[!htb]
\centering
\includegraphics[width=\hsize]{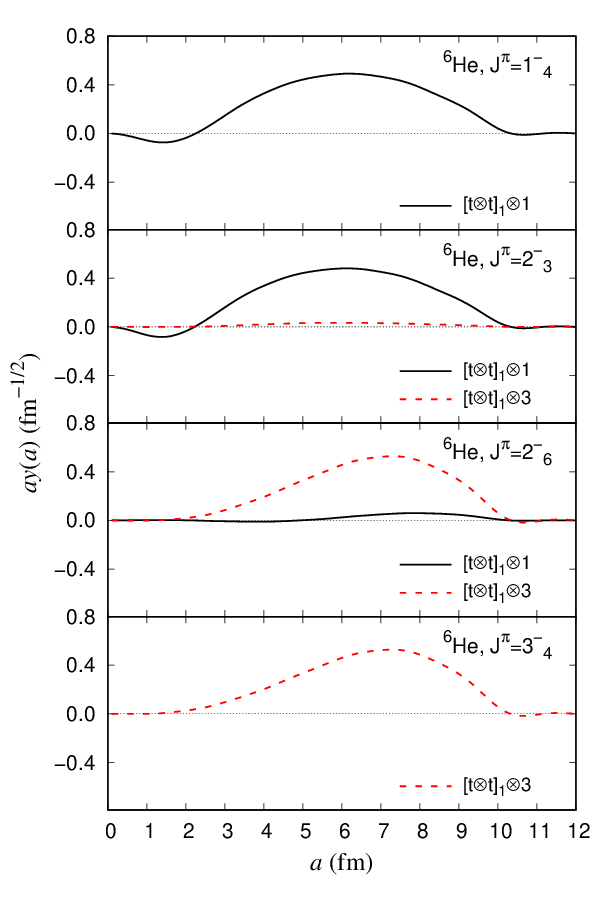}
\caption{The calculated RWA of $t+t$ configuration in negative-parity resonant states with prominent $t+t$ structure.}
\label{fig5}
\end{figure}

% energy level - calculation details
For determining the $t+t$ resonances of $^6\mathrm{He}$ above the threshold, we perform the GCM calculation with a $t+t$ cluster model.
The radius constraint method (RCM)~\cite{YFunaki} is applied to remove the continuum components from the obtained resonant states.
However, we found that the energies of these resonances of $t+t$ show a strong dependence on the radius cutoff $R_\mathrm{cut}$.
In the present calculation, we adopt $R_\mathrm{cut}=5.0~\mathrm{fm}$.
Besides, it is difficult to describe well the ground state $0_1^+$ and the first excited state $2^+_1$ of $^6\mathrm{He}$ considering merely $t+t$ configuration, as a result that these states consist of significant components of $\alpha+n+n$ clustering configurations.
Consequently, we lower the energies of the ground and the first excited states by adding a few $\alpha+n+n$ configurations, in which the neutrons are close to the $\alpha$ cluster (shell-model-like) to avoid inducing continuum states.

% energy&RWA - results description
The energy spectra are shown in FIG.~\ref{fig3}.
On the left is the energy spectrum obtained by the $t+t$ cluster model.
Despite the too-high ground state and first excited state, several states above the $t+t$ threshold show good agreement with the experiment results.
In the middle columns, we display the result obtained by coupling the $t+t$ and shell-model-like $\alpha+n+n$ configurations.
The introduction of $\alpha+n+n$ configurations significantly lowers the ground state energy as well as the first excited state energy, making them consistent with the experimental values.
More importantly, we notice that some of the states obtained from the $t+t$ cluster model are quite insensitive to the introduction of $\alpha+n+n$ configurations.
We speculate that these states should be well-developed $t+t$ resonance states, or contain considerable $t+t$ structure components. 
By evaluating the $t+t$ RWA for states in both the calculated spectra, we recognize the $t+t$ resonance states in the results obtained by the $(t+t)\oplus(\alpha+n+n)$ model (red lines), which exhibit nearly identical RWA curves with the corresponding states obtained by $t+t$ model.

Fig.~\ref{fig4} shows the $t+t$ RWA results for the ground state, the first excited state, and two positive-parity $t+t$ resonances, whereas Fig.~\ref{fig5} shows the results for negative-parity $t+t$ resonances.
The RWA has considerable amplitudes for the ground and the first excited states when the inter-cluster distance $a$ is small and decreases rapidly for larger $a$.
In contrast, the $t+t$ resonant states are featured with extended amplitudes. 
As shown in Eq.~(\ref{eq:rwa}), the RWA is defined as the overlap amplitude between the wave function of $^6\mathrm{He}$ and the coupling wave functions of two $t$ clusters.
As a result, the absolute value of RWA reflects the probability of cluster formation at a given distance $a$.

In the GCM analysis of the ${^3\mathrm{He}}+{^3\mathrm{He}}$ scattering system~\cite{Baye1981}, the inter-cluster orbital angular momentum was identified as a critical quantity determining state characteristics, due to spin-coupling effects inherent to fermion systems.
In the present study, through RWA results, we identify prominent angular momentum channels across various states.
For positive-parity states, the triton clusters form spin-singlet states with $s$-wave and $d$-wave relative motion, respectively.
Conversely, the negative-parity states exhibit spin-triplet states, where $p$-wave dominates in the $1^-$ and the first $2^-$ states, and $f$-wave dominates in the $3^-$ and the second $2^-$ states.

The current calculation provides a robust and consistent description with the coupling configurations of \(\alpha + n + n\) and \(t + t\), not only characterize the system at the lowest energy levels but also offer an accurate description of the \(t + t\) states above the threshold, highlighting their relevance in the broader spectrum. 

The GCM calculations indicate that the $t+t$ cluster resonant states above the threshold exhibit a rich variety of quantum state characteristics. This richness arises primarily from the fermionic nature of the $t$ cluster, resulting in significantly more quantum energy levels for the $t+t$ resonance compared to the $\alpha+\alpha$ resonance in $^{8}$Be. Due to the lack of precise measurements and the inability to extract the spin-parity of the observed resonant state, it is impossible to confirm which quantum state the observed energy level corresponds to. Consequently, effectively comparing the energy levels measured in different experiments becomes challenging. Furthermore, multiple quantum states are permissible for $t+t$ resonance near the threshold, with experimental measurements concentrated in this region, as indicated by the black-colored columns in FIG. ~\ref{fig3}. Therefore, accurately measuring the $t+t$ resonant state in experiments is essential.

\section{Conclusions}
The $^{6}$He isotope is a neutron-rich nucleus that exhibits above-threshold $t$+$t$ cluster states, providing valuable insights into the behaviour of loosely bound nuclear systems and their formation mechanisms. The discovery of multiple resonant states is particularly significant as it contrasts with the traditional $\alpha$ cluster structures observed in other light nuclei, highlighting the unique fermionic properties of the triton cluster. The excitation energy spectrum above the $t$+$t$ threshold in $^{6}$He is measured during the decay of $^{9}$Li on a $^{208}$Pb target at 32.7 MeV/nucleon. The resonant states are reconstructed using the invariant mass method and reveal peaks at $13.9 \pm 0.3$ and $15.0 \pm 0.3$ MeV. These results are compared with previous observations, although inconsistencies remain. Microscopic cluster model calculations, including the coupling between $t+t$ and $\alpha+n+n$ configurations, are carried out to explore the $t+t$ resonant states in $^{6}$He. The theoretical energy spectra are in good agreement with the experimental results. To further investigate the clustering structure of the theoretical states, we calculate the RWA of the $t+t$ channels. As expected, the identified $t+t$ resonant states show significant amplitudes in the $t+t$ RWA, reinforcing their clustering nature. This study highlights the complexity of the energy level discrepancies in the excited states of $^{6}$He and provides accurate theoretical predictions. These results underline the need for more accurate and comprehensive experimental studies to further explore the structure of the $t+t$ resonance in $^{6}$He. A future experimental setup should feature high detection efficiency, high statistics, and high precision to accurately determine resonant states with spin-parity. Angular correlation analysis is a valid method for experimentally identifying spin-parity for the sequential breakup reaction described as $a(A, B^{*} \rightarrow c + C)b$ ~\cite{Freer1,Finkel}.

The identification of $t+t$ resonances in $^{6}$He provides critical insights into fermionic clustering, distinguishing it from traditional bosonic $\alpha$-clustering and advancing our understanding of exotic nuclear structures. These findings help refine theoretical nuclear models by addressing discrepancies in observed excitation spectra and guiding improvements in microscopic cluster calculations. Additionally, exploring these fermionic cluster systems and their potential to form quasi-molecular states upon adding further nucleons contributes to broader studies of neutron-rich matter, with implications for nuclear astrophysics and the structure of drip-line nuclei.

\section*{Acknowledgements}
This work is supported by the National Key R$\&$D Program of China (2023YFA1606404，2023YFA1606701，2022YFA1602402). This work is supported in part by the National Natural Science Foundation of China under contract Nos. 12175042, 11890710, 11890714, 12047514, 12147101, 12475138, 11875297, and 12347106, the Guangdong Major Project of Basic and Applied Basic Research No. 2020B0301030008, the STCSM under Grant No. 23590780100, and the Natural Science Foundation of Shanghai under Grant No. 23JC1400200.

\end{CJK*}
\end{document}